\documentstyle[aps,prb,epsfig]{revtex}
\draft
\begin{document}
\title{Total energy differences between SiC polytypes revisited}
\author{Sukit Limpijumnong and  Walter R. L. Lambrecht}
\address{Department of Physics, Case Western Reserve University, 
Cleveland, OH 44106-7079}
\date{\today}
\maketitle
\begin{abstract}
The total energy differences between various SiC polytypes
(3C, 6H, 4H, 2H, 15R and 9R) were calculated using the full-potential
linear muffin-tin orbital method using the 
Perdew-Wang-(91)  generalized gradient
approximation to the exchange-correlation functional 
in the density functional method.
Numerical convergence versus {\bf k}-point sampling 
and basis set  completeness are demonstrated to be better than 1 meV/atom. 
The parameters of several generalized 
anisotropic next-nearest-neighbor Ising models are 
extracted and their significance and consequences for epitaxial growth 
are discussed. 
\end{abstract}

\section{Introduction}
Despite many years of study the origin of polytypism in
SiC is still not completely understood.  
A much debated question is whether polytypism is a manifestation of 
kinetic factors during growth or whether polytypes should be viewed
as distinct (possibly metastable) thermodynamic phases with 
a specific stability range of external parameters (such 
as pressure, temperature).
In a thermodynamic approach to the problem, the  most important quantities 
are the total free energy differences between the various polytypes. 
A major contribution to the latter is the energy difference 
at zero temperature.
Vibrational entropy contributions at higher temperature were discussed by 
Heine et al.\cite{Heine,Cheng90a} and  Zywietz et al.\cite{Zywietz}  
Several groups have performed first-principles local density
functional calculations of these energy 
differences using the norm-conserving pseudopotential plane wave  
method.\cite{Cheng88,Cheng90b,Park,Kackell,Karch} However, there are
significant discrepancies between the results of various 
calculations for these energy differences, which 
are of order of a few meV/atom or less. More seriously, the three more
recent calculations appear to invalidate some of the 
important conclusions drawn from these calculations by the early work 
of Heine et al.\cite{Heine}

Heine et al. \cite{Heine} discussed the relative energy of 
polytypes in terms of a generalized anisotropic next-nearest-neighbor
Ising (ANNNI) spin model
in which the energy of a given polytype (per atom) is written as
\begin{equation}
E=E_0-\frac{1}{N}\sum_{i,n} J_n \sigma_i\sigma_{i+n},
\end{equation}
in which $N$ is the number of layers in the system, a ``spin'' 
$\sigma_i=\pm1$ is associated with each (close packed) SiC double layer 
such that parallel spins represent a locally cubic 
stacking and antiparallel spins represent
a locally hexagonal stacking. The parameters $J_n$ 
represent the interlayer interaction between succesively farther 
removed layers and $E_0$ is a common energy reference. 
In terms of this model truncated beyond $n=3$ 
the energies of some of the polytypes of interest 
are given in column 2 of Table \ref{t-annni}.    
According to Heine et al.\cite{Heine},  what distinguishes SiC from other
semiconductors, and leads to the multitude of stacking arrangements 
constituting polytypism, is that $J_1\cong-2J_2>0$ 
(with $J_n\ll J_{1,2}$ for $n>2$).
For this special ratio of $J_1/J_2$, a multi-phase degeneracy point occurs in
the ANNNI model corresponding to all phases consisting of successive
bands of 2 or 3 parallel spins (which in the following we will call
``2-3 banded'' polytypes). This would explain
the relatively frequent occurence of polytypes
such as 6H (which is $\langle3\rangle$ in Zhdanov notation \cite{Zhdanov}
indicating that it consists of bands of 3 parallel spins), 4H  or
$\langle2\rangle$, 15R or $\langle32\rangle$,  in contrast 
to polytypes such as 8H or $\langle4\rangle$, 10H  or $\langle5\rangle$, 
and 2H  or $\langle1\rangle$ which are rather rare. 
Surprisingly, recent calculations \cite{Park,Kackell,Karch} 
found that  $J_1<|J_2|$, a condition  very far away from 
the multi phase degeneracy point.
Secondly, in some of these results,\cite{Park,Kackell}
the energy energy difference $E_{2H}-E_{3C}$  is found
to be smaller than the energy difference 
between 3C and any of the other low energy polytypes. This makes it 
difficult to understand why 2H is such a rare polytype. 

The purpose of the present paper is to systematically re-evaluate these
energy differences of polytypes including some new ones
and to discuss the meaning of the ANNNI model parameters in the light
of these results. 
Since the accuracy is a crucial matter here, we next discuss the
computational method and associated convergence parameters in some detail.

\section{Computational method and convergence tests}
The computational method employed is the full-potential linear muffin-tin
orbital method as implemented by Methfessel\cite{Methfessel}  
and van Schilfgaarde\cite{vanschilf}. The total energy  is calculated 
using the density functional method using the 
generalized gradient approximation
(GGA) for the exchange-correlation energy of Perdew and Wang (PW).\cite{PW91}
For the $E_{2H}-E_{3C}$ energy difference, which is of particular concern
below, we verified that other 
exchange-correlation functionals, such as the Langreth-Mehl GGA \cite{Langreth}
and the Ceperley-Alder \cite{Ceperley} and Hedin-Lundqvist \cite{Hedin}
parametrizations of the local density approximation yield results which do 
not differ by more than 1 meV/atom from those for the PW-GGA 
adopted in the rest of this paper.   

In all results presented below, we used the ideal 
structures but relaxed the total energies with respect to volume.
All polytypes were found to closely obey the expected relation 
$a_h=a_c/\sqrt{2}$ and $a_c$ was found to be 4.33 \AA, within 1 \% of the 
experimental value. To check the uncertainties introduced by using ideal 
structures, we performed relaxations for 2H-SiC.
We found $c/a=1.644$, i.e. slightly larger than the ideal ratio
$c/a=\sqrt{8/3}=1.633$ in good agreement with experiment,\cite{Wyckoff}
which gives $c/a=1.641$. 
We obtain $u=0.3745$ which is very close to the ideal value of $3/8$.
The important point is that the total energy per atom in 2H 
was reduced by only 0.6 meV/atom  by relaxation of the structure. 
In the above calculation, an intra cell parameter $u$ relaxation
was performed for each $c/a$.
This energy lowering 
is consistent with the value estimated from the elastic constants
for a distortion from the minimum energy $c/a=\eta$ to the ideal 
$c/a$, given by 
$\Delta E= (1/9) (\delta \eta/\eta)^2\Omega [C_{33}-2C_{13}
+(C_{11}+C_{12})/2]$, in which $\Omega$ is the volume per 
Si-C pair. This expression equals 0.56 meV/atom using the elastic constants
given in Ref. \onlinecite{Lamelast}. 
Since  other polytypes of type $2nH$ are found experimentally to  have 
$c/a$ values closer to the ideal value of $n\sqrt{8/3}$, 2H is the 
extreme case and places an upper limit on the errors 
introduced by using ideal structures. 
The effect of internal cell structural parameters is even an order 
of magnitude lower. 
From the TO-phonon frequency in SiC of 23 THz, \cite{Lamelast} 
we can estimate the force constant $k$ for the Si-C bondlength distortions
to be 18 eV/\AA$^2$. Thus the change in energy per bond related to a bond-lenth
change of $\delta l=(\delta u) c$ is $\Delta E=(1/2) k (\delta l)^2$.
This gives only 0.03 meV/atom for our calculated $\delta u$ in good agreement
with our direct calculation. High precision theoretical determinations 
of the atomic relaxations were reported by K\"ackel et al.\cite{Kackell}
They confirm that the bond lengths differ by less than 0.3 \% from the 
ideal bond length which according to the above estimate would give at most 
0.14 meV/atom for the relaxation energy.
As far as atomic relaxation effects on 
the total energies is concerned, our results 
differ substantially from theirs.  In their results without atomic relaxations
or cell-shape relaxations, 2H lies about 8 meV/atom  
above 3C and the energy of the polytypes
increases  monotonically with hexagonality. They find the 
internal cell atomic relaxations 
to have a marked effect on all hexagonal polytypes resulting 
in a lower energy than 3C for 6H and 4H  and a substantial reduction 
of the 2H to 3C energy difference to only 1 meV/atom. This implies
that the relaxations would produce 
relative changes  of 3, 4, and 7 meV/atom for
6H, 4H and 2H respectively. This is inconsistent with the above estimates 
based on elastic and force constants and with our explicit calculations. 
The origin of this discrepancy is not entirely clear.
However, the comparison between 
their relaxed and unrelaxed energies is complicated by their use of different
${\bf k}$ point sets for the two calculations.
As will be shown below, converged Brillouin zone
integrations are an important requirement for drawing conclusions about
polytype energy differences.
As we will show below, we find 6H and 4H to have lower
energy than 3C without cell shape or internal position 
relaxations.

\begin{figure}
\begin{center}
\mbox{\epsfig{file=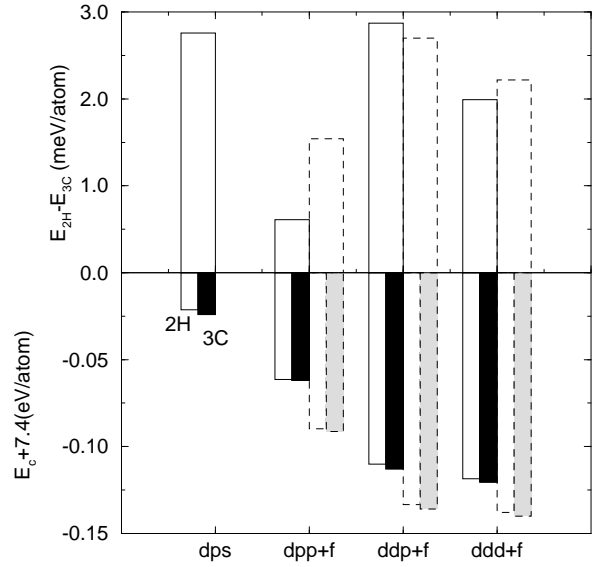,height=8cm}}
\caption{Basis set convergence of $E_{2H}-E_{3C}$. Bottom graph 
shows cohesive energies of 2H and 3C as open and filled bar graphs
with various basis sets as indicated. Top graph shows $E_{2H}-E_{3C}$.
Dashed  (full) lines are results for which $f$-orbitals are (not) 
included.}\label{fbset}
\end{center}
\end{figure}

With respect to self-consistency,
all total energies were converged to better than 0.1 meV/atom.
The contributions to the total charge density from each angular  momentum
component were converged to a root mean square error less 
than 10$^{-4}$ electron.
Within FP-LMTO, the wave functions are expanded in an extended 
basis set of muffin-tin orbitals with different spatial decay constants 
(i.e spherical Hankel envelope function exponents $\kappa$). 
Fig. \ref{fbset} shows results for different basis sets for 
the $E_{2H}-E_{3C}$ energy difference and the individual cohesive energies
of $E_{2H}$ and $E_{3C}$. The notation for the basis set is illustrated 
as follows:
$dps$ means up to $d$ orbitals for the first $\kappa=-0.05$ Ry, up to $p$ for 
second $\kappa=-1$ Ry and one  $s$ orbital for the third $\kappa=-2.3$ Ry.
The unfilled bars in the bottom graph give $E_{2H}$, the filled ones $E_{3C}$.
The top graph gives their energy difference in meV/atom. The dashed lines
indicate the corresponding information for the same basis sets with 
$f$-orbitals added for the first $\kappa$. 
We can see that the contributions of each orbital to the total cohesive
energies are several 10 meV and that increasing 
the basis set decreases the energy. 
Adding the $f$-orbitals makes about a $-$30 meV contribution  
independently of which basis set they are added to. 
The third $\kappa$ $d$-orbital contributes only about $-7$ meV to the 
total energy. The most important point is that the polytype energy difference
is stable at 2.4$\pm0.3$ meV/atom for the four most complete basis sets
considered.  Adding empty sphere orbitals $s$ and $p$ and second $\kappa$ $s$ 
to the $ddp$ basis set changed the energies by only $-8$ meV
and is thus also considered ineffective. 
For polytypes with many atoms per unit cell, the calculations with the 
basis sets larger than $fdp$  tend to become unstable.
If the basis set is very close to completeness, slight numerical 
errors can make the basis set appear to be overcomplete or 
linearly dependent. 
The optimal basis set is thus considered to be $fdp$ and used systematically
for the other polytypes.

The integrations over the interstitial region are done using an 
auxiliary set of spherical Hankel functions times spherical harmonics
for expansion of products of two Hankel functions. These expansions
are cut-off at $l_{max}=6$. We found that this cut-off 
is necessary to make the results stable and independent of the sphere
radii choice. The empty spheres were chosen to be nearly touching
with two empty spheres equal in size to the atomic spheres (Si and C
being chosen equal) in each cubic stacking double layer unit and 
a large (1.134 $s_{atom}$) and small empty sphere (0.666 $s_{atom}$) 
in each hexagonal unit.  The large spheres
occupy the empty channel in the wurtzite structure. 
That is, if atoms are taken to sit in  $A$ and $B$ positions in 
the basal plane, the large empty spheres occupy the $C$ positions 
in the plane at a height halfway between the bonding Si and C
atoms in the $A$ position. The small spheres
occupy the sites halfway between the Si and C atoms opposite
to the nearest neighbor Si-C bond along the c-axis. 
In cubic SiC, the spheres occupy about 68 \% of the unit cell volume.
In 2H they occupy 63 \% of the volume and in other polytypes the 
filling is in between these values in proportion to the 
degree of hexagonality (i.e. the ratio of the 
number of hexagonally stacked layers $h$ to the total 
(i.e hexagonal and cubic $c$)
number of layers  $h/(h+c)$).

\begin{figure}
\begin{center}
\mbox{\epsfig{file=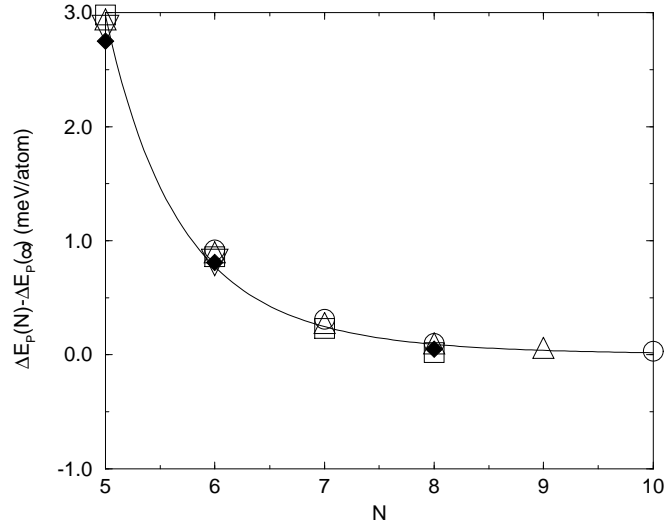,height=8cm}}
\caption{Brillouin zone sampling convergence. $N$ is the number
of divisions along the two basal plane reciprocal lattice vectors.
Circles: 2H, squares: 4H, filled diamonds 6H, upward triangles 9R, downward
triangle 15R. The full line curve is a powerlaw fit $e^{13}N^{-7.4}$. 
The values of $\Delta E_P(\infty)$ are given in Table \ref{t-annni}.}
\label{fkcon}
\end{center}
\end{figure}

The next convergence issue to consider is the Brillouin zone integration.
The Monkhorst-Pack \cite{Monkhorst} special {\bf k}-points sampling 
technique is used with the number of divisions along reciprocal lattice
vectors in the basal plane equal to $N$ and along the $c$-axis equal to $M$.
For longer polytypes (along the c-axis), one needs fewer divisions 
$M$ along the c-axis. Rather than picking exactly equivalent sets 
for each polytype, and thus counting on error cancellation, 
we picked $M$ large enough to ensure absolute convergence. For 
2H, we used $M=N$ and for longer polytypes we reduced to 
$M=N/2$ for the larger $N$ values.
Fig. \ref{fkcon} shows the results for various 
polytypes as a function of $N$.
The quantity shown is $\Delta E_P(N)-\Delta E_P(\infty)$,
where $\Delta E_P(N)=E_P(N)-E_{3C}$, the energy difference 
for a given polytype $P$ from the 
absolutely converged value of $E_{3C}$ calculated with $N=M=10$,
and the value of $E_P(\infty)$ is estimated by extrapolation 
so as to ensure that all results fall on a universal curve.
This clearly shows that the final values $\Delta E_P(\infty)$ 
are converged to better than 0.5 meV/atom. 

\section{Results and Discussion}
\begin{figure}
\begin{center}
\mbox{\epsfig{file=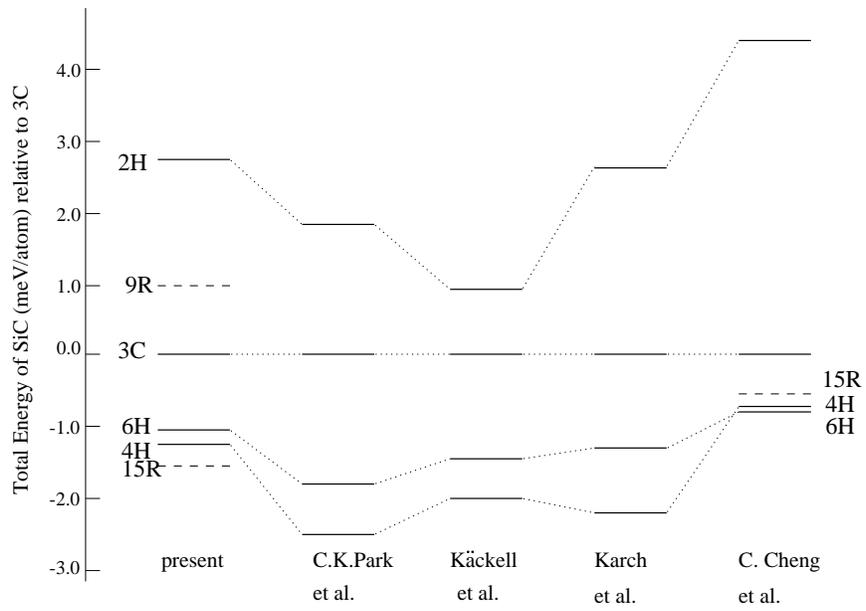,height=8cm}}
\caption{Energy differences between various polytypes: comparison
with other calculations.}\label{fcomp}
\end{center}
\end{figure}

The converged energy differences of the polytypes 
with respect to 3C, i.e. $\Delta E_P(\infty)$ 
as defined in the previous section,
are given in column 3 of Table \ref{t-annni}. 
They are compared to those of previous calculations
in the literature in Fig. \ref{fcomp}. 

Next, we extract the $J_n$ parameters. Columns 4 and 5 
correspond respectively to truncation at $n_{max}=2$ and $n_{max}=3$ using the 
energy differences $E_{2H}-E_{3C}$, $E_{4H}-E_{3C}$ as input  in the 
first case  
and additonally $E_{6H}-E_{3C}$ in the latter case. 
The other polytypes then allow for a check 
of the consistency of this model. 
We find the $J_n$ parameters $J_1$ and $J_2$, as listed in Table 
\ref{t-annni}, to be nearly independent of whether or not $J_3$ 
is included. Furthermore, we find $J_1>|J_2|$. 

Our results are somewhat closer to  those of Heine et al.\cite{Heine} 
than  the other recent results, particularly that 
$E_{2H}$ is higher above $E_{3C}$ by an amount significantly larger 
than the other polytype energy differences. 
Also, we find the various 2-3 banded polytypes 
to be closer to each other than 
in the other calculations.  Nevertheless, our results  are far from the 
multiphase degeracy point $J_1=-2J_2$.
As Heine pointed out, the energy of a twin boundary, i.e. the 
energy cost of a boundary between all up-spin all down-spin 
cubic half crystals is given by 
\begin{equation} 
E_{twin}=2(J_1+2J_2).
\end{equation}
According to Heine et al., this is nearly zero and hence explains
why many twin boundaries in an otherwise cubically stacked crystal
are likely to occur. 
With our present values of the $J_n$ parameters,
the energy cost of a twin is {\it negative}. 
This implies that twins are even more favorable than in Heine's model.
Hence, there is no contradiction at all with the observation 
of a predominance of 2-3 banded polytypes.

Consistently with other recent work we find 4H to have lower energy than 6H.
In the FP-LMTO calculations, we find 15R as lowest energy polytype.
In the ANNNI model we find 15R to lie in between 4H and 6H with 4H 
the lowest energy polytype. This is slightly more expected since 15R 
is intermediate  in character between 4H and 6H.
This discrepancy, which is smaller than 0.5 meV, may 
be beyond the accuracy of our FP-LMTO calculations in view of the fact
that the computational convergence is most challenging for the largest
polytype. 
As expected, the hypothetical\cite{9R,icscrm} 9R polytype 
with a high degree of hexagonality
(66\%) is found to have higher energy than 3C  but lower than 2H.

The ANNNI model appears to somewhat underestimate the  energy
of 9R. This suggests that other terms in the effective Hamiltonian may 
be required. A term 
\begin{equation}
K\sum_i\sigma_i\sigma_{i+1}\sigma_{i+2},\sigma_{i+3},
\end{equation}
was suggested by Cheng et al.\cite{Cheng88} 
The additional energy for each polytype 
due to this term is given in column 5 in Table \ref{t-annni}. 
Column 6 shows that this term allows us to fit 9R exactly without 
affecting  the energy of 15R significantly.

We next consider the predictions of the model for a few other polytypes. 
Another polytype of high hexagonality (80 \%) was recently considered
\cite{icscrm} and 
labeled 15R' or $\langle1112\rangle$.  
Its energy within the ANNNI model is given in the bottom 
section of Table \ref{t-annni}. As expected, it is higher in energy than 
9R but still lower than 2H. We do not interpret this as 
an indication that these particular
periodic stacking arrangements are more likely (because they seem excessively
complicated) but rather as an indication that a high density of stacking faults
is likely to occur in 2H.
For any $2nH$ polytype with $n\ge3$ the energy 
difference from 3C can be 
written as $\frac{2}{n}(J_1+2J_2+3J_3-2K)$. This shows that for 
$n\rightarrow\infty$, it will approach zero as expected since 3C corresponds
to $\infty H$, but only very slowly. In fact, the energies of 8H and 10H 
are seen in Table \ref{t-annni} to be still rather close 
to those of the 2-3 banded polytypes, consistent with 
the fact that these polytypes have indeed been observed.

As for the phonon contributions to the free energy, (here denoted $F_P$) 
we note that Heine et al. \cite{Heine} 
obtain  a result which is the opposite of 
that found by Zywietz et al.\cite{Zywietz}, namely 
$F_{4H}>F_{6H}$ and increasing with temperature. This tends 
to stabilize 6H at high temperatures whereas Zywietz et al.\cite{Zywietz}
find 4H to become even more stabilized at higher than at lower
temperatures without affecting the polytype free energy ordering.
We note that with our calculated $E_{4H}-E_{6H}$ at zero temperature,
and Heine's values for the phonon contribution, the transition 
from 4H stability to 6H stability is predicted to occur above 8000 K,
i.e. well above the melting temperature of SiC. With Zywietz et al.'s
phonon contributions, no stabilization of 6H will ever occur.
We conclude that either way, there is no substantial evidence from the 
calculations that the polytypes would have a well-defined temperature
stability region. We think it is much more likely that the slightly 
different tendencies  for 4H and 6H growth in dependence on the 
growth temperature is due to kinetic factors. In fact, these experimental
tendencies have not  unequivocally  been established.

Heine et al.\cite{Heinecubic} also argued that the 3C dominance in epitaxial
growth could be explained by assuming that only the surface
layer stacking is determined by the equilibrium energy condition
but that the stacking is not subsequently 
re-arranged after the layer is burried into the growing crystal.
Since the energy difference for adding one surface layer 
to a substrate with opposite spin of the top layer as  
opposed to equal spin is $J_s=2(J_1\pm J_2)$, 
with $\pm$ depending on whether the 
next layer down has equal or opposite spin, cubic stacking is 
always favored as long as 
$J_1+J_2>0$. As in Heine et al.' s results, 
and in contrast to other recent results,\cite{Park,Kackell,Karch}
our present results satisfy this requirement although only barely so.
Of course, we caution  that  these interlayer 
interactions may change at a surface.  If $J_1+J_2<0$, 
on the other hand, a 4H stacking would always be preferred as can 
easily be checked by following the same argument as given by Heine.
The point is that second layer interactions, which are ``antiferromagnetic''
are then dominant. Thus, if we start from two equal spins in the top layers,
the next growing layer must have opposite spin. The new surface ends then 
in two opposite spins and the following layer must  have the same 
spin as the one last deposited, after which the cycle repeats.
This is inconsistent with experimental observations.
Independent nucleations on large terraces tend to have the 3C structure,
which usually is accompanied by a large amount of so-called double
positioning boundaries.

Given that the preference for cubic stacking during growth is 
so small, the question arises whether this is really relevant. 
To address this question, we must consider size effects of the growing 
fragments. For a 2D  island of $N_i$ spins (or SiC units), 
the energy differences for 
being in a cubic or hexagonal stacking on top of a substrate
should be of order $N_i J_s$.  This implies that up to $N_i J_s\approx k_BT_G$ 
with $T_G$ the growth temperature 
and $k_B$ Boltzman's constant, or for a typical growth temperature of 1500K,
and using $J_s=0.3$ meV/SiC unit, up to $N_i\le600$, 
there should be virtually no distinction in energy
between either stacking. On the other hand, islands will definitely
tend to be of a well-defined spin. This is because a lateral 
spin-boundary corresponds ultimately to a defect such as an incoherent
twin boundary. The energy of the latter is typically of the order of 
several eV/atom.\cite{Kohyama} This is because there are serious
disruptions of the tetrahedral bonding associated with such boundaries,
including wrong bonds (C-C or Si-Si) and  possibly dangling bonds. 
Thus, atoms migrating on the surface will have  a strong tendency to 
adjust their spin  (i.e. stacking with respect to the underlying layers)
to that of the growing island to which they are attaching. This explains
why well-defined polytype structures can evolve 
even if the growth does not occur in a strict layer-by-layer fashion
in spite of the energy differences for different stacking for each atom
being much smaller than the growth temperature. 
Only for islands of the above defined size, which corresponds to $\sim$10 nm
in diameter, one expects that the interactions with underlying layers
become relevant. A predominace of cubic stackings with respect to the 
underlying layers assumes that such 2D islands can still adjust 
their stacking position by moving as a whole. Although this might
seem to require overcoming a significant energy barrier, motion of 
islands might occur by a 2D dislocation motion.   
In the above estimate, we used $J_s=2(J_1+J_2)$  neglecting $J_3$ 
and $K$ interactions. We also assumed growth on a cubic substrate
and renormalized to energies per SiC unit rather than per atom.
For growth on other polytype substrates or when including $J_3$ and/or $K$,
the interaction $J_s$ becomes somewhat larger and hence the critical 
island size somewhat smaller, but the general argument does not change.
Even though a preference for cubic stacking can thus be rationalized, 
a certain number of 
double positioning boundaries are expected because 
some islands of opposite spin may become trapped in an initially
unfavorable stacking due to the randomness of the initial nucleation events.
A step-flow growth mechanism seems to be the only plausible mechanism
for stabilizing other polytypes during epitaxial growth and depends 
crucially on the sizes of the terraces and the surface diffusion (hence
growth temperature).\cite{Matsunami}

\section{Conclusions}
In conclusion, we have carefully re-evaluated the zero-temperature
energy differences between polytypes of SiC using well-converged all-electron 
density functional calculations. We find that 
the ANNNI model with up to second nearest neighbor layer interactions 
already provides a good description of the polytype energy differences 
with shlight improvements beging obtained 
by including a third layer interaction and 
a 4-spin term. Even though the values for $J_1$, $J_2$ 
do not correspond to the multi-phase degeneracy
point, the predominance of polytypes of narrow bands of
cubic stacking (typically 2-3 banded) can readily be explained by the 
fact that $J_1>0$ and  the twin boundary energy cost is negative. 
Our results agree closer with Heine et al.'s \cite{Heine} work 
than other recent 
calculations, in the sense that we obtain $J_1>|J_2|$, the 2-3 banded 
polytype energies closer to each other and the 2H energy 
significantly higher than that of 
3C. We stress that this is not due to our neglect of relaxations because 
the latter were shown to be at most 0.6 meV/atom.
We nevertheless find the energies of 4H and 6H  to differ substantially
enough to preclude a well-defined temperature stability region 
for each polytype when using literaure data for  the vibrational 
free energy contributions. This suggests that polytypes are  kinetically 
determined metastable phases rather than true thermodynanic phases. 
Some consequences for epitaxial growth were discussed.   In particular,
we extended Heine et al.'s arguments concerning the tendency for 3C growth
to occur if only equilibrium  of the top-surface layer is required
by considering the island size effects.
We also showed that for $J_1<|J_2|$,
4H would always be stabilized, which is inconsistent with experiment.

We thank B. Segall for useful discussions.
This work was supported by NSF DMR-95-29376.

\begin{table}
\caption{Energy difference $\Delta E_P=E_P-E_{3C}$ for various polytypes P
in meV/atom}
\begin{tabular}{lcccccc}
P  & ANNNI        & FP-LMTO &   
$n_{max}=2$ \tablenote{Using $J_1=1.350$, $J_2=-1.285$, $J_3=0$, extracted from 
first two polytypes} & $n_{max}=3$ \tablenote{Using $J_1=1.528$, 
$J_2=-1.285$, $J_3=-0.177$ meV/atom, extracted from first three polytypes}
& $K$-term & $n_{max}=3$, $+K$ \tablenote{Using 
$J_1=1.781$, $J_2=-1.275$, $J_3=-0.431$, $K=-0.244$ meV/atom extracted
from first four polytypes}  \\ \tableline 
2H & $2J_1+2J_3$ & 2.7 &    2.7     &  2.7 & 0 &  2.7 \\
4H & $J_1+2J_2+J_3$ & -1.2 & -1.2   &  -1.2 & 0 & -1.2 \\  
6H & $\frac{2}{3}J_1+\frac{4}{3}J_2+2J_3$ & -1.05 & -0.08 & -1.05 & 
$-\frac{4}{3}K$ & -1.05 \\
9R & $\frac{4}{3}(J_1+J_2)$ & 1.0 & 0.1 & 0.3 & $-\frac{4}{3}K$ & 1.0 \\
15R & $\frac{4}{5}(J_1+2J_2+2J_3)$ & -1.5 & -1.0 & -1.1 & $-\frac{4}{5}K$ & 
-1.1 \\ \tableline
15R' & $\frac{4}{5}(2J_1+J_2+J_3)$ & & 1.1 & 1.3 & $-\frac{4}{5}K$ & 1.6 \\
8H   & $\frac{1}{2}(J_1+2J_2+3J_3)$& & -0.6 & -0.8 & $-K$ & -0.9 \\
10H  & $\frac{2}{5}(J_1+2J_2+3J_3)$& & -0.5 & -0.6 & $-\frac{4}{5}K$ & -0.7 \\
\end{tabular}\label{t-annni}
\end{table}

\end{document}